\documentclass{ptptex}
\usepackage{graphicx}

\inst{Department of Earth and Planetary Sciences, Kobe University, Kobe 657-8501, Japan
\begin{center}
(Received February 26, 2005)
\end{center}}

\abst{Subramanian, Pujari and Becker (2004) claim that the correct expression 
for the angular momentum transport in an accretion disc, which is 
proportional to \(d \Omega/d R\), can be derived on the basis of the 
analysis of the epicyclic motion of gas parcels in adjacent eddies 
in the disc.  We study their argument and show that their 
derivation contains several fundamental errors:
1) the biased choice of the desired formula from an infinite number
 of formulae;
2) the biased choice of parcel trajectories;
 and 3) confusion regarding the reference frames. 
Following 1) we could derive, for example, a
(invalid) formula in which the angular momentum transport is proportional
to $d v_{\phi}/dR$, and from 2) we could even prove that the angular momentum 
transport is either inward or null. We present the
 correct approach to the problem
of angular momentum transport in an accretion disc in terms of mean free path
theory.}

\input{tcilatex}

\begin{document}

\title{Comments on ``Angular Momentum Transport in Quasi-Keplerian Accretion
Disks'' }
\author{Eiji \textsc{Hayashi}, \mbox{Hiromu \textsc{Isaka}} 
\mbox{
and Takuya \textsc{Matsuda}} }
\maketitle

\section{Introduction}

Elucidating the mechanism(s) involved in the transport of angular momentum 
in accretion
discs is a long-standing problem. Observations show that the
gas in an accretion disc rotating around a compact object 
in a close binary does
gradually accrete on the compact object. This can only happen if the angular
momenta of the molecules/fluid parcels constituting the rotating gas are
transported outwardly through the disc. (In the present paper we do not
distinguish between molecules in kinetic theory and fluid parcels in
turbulent motion, unless explicitly stated.)  In recent years, the theory
of angular momentum transport by magnetically induced turbulence is
getting fashionable (e.g. Balbus and Hawley\cite{B-H}).  Nevertheless, it
is important to derive a correct viscosity formula in a rotating gas, and
we concentrate on the viscosity due to either molecular motion or 
hydrodynamic turbulence in the present paper.

Consider a rotating gas flow represented by the velocity field $\mbox
{\boldmath$u$}
=(0, R\Omega(R))$ in cylindrical coordinates, where $R$ and $\Omega$ are the
radial distance and angular velocity, respectively. Then the correct $R-\phi$
component of the viscous stress should be

\begin{equation}
\sigma_{R\phi} = - \eta R(d\Omega/dR),  \label{circular}
\end{equation}
where $\eta$ is the viscosity coefficient.

In order to prove that the viscous stress is proportional to $d\Omega /dR$,
some textbooks apply a simple mean free path theory, in which a gas
molecule/parcel jumps over a distance equal to 
the mean free path, carrying with it the angular
momentum it possessed at the place where 
it originated (Frank, King and Raine\cite%
{F-K-R}, hereinafter referred to as FKR; Hartmann\cite{H}).
However, Hayashi and Matsuda\cite{H-M} (hereinafter HM) pointed out that if
the proof is carried out correctly, it should lead to the incorrect formula $%
\sigma _{R\phi }\propto -d(R^{2}\Omega )/dR$ (see also Subramanian, Pujari
and Becker\cite{S-P-B}, hereinafter referred to as SPB). This is clear when
one applies a simple mean free path theory (see, for example, Vincenti and
Kruger\cite{V-K}).  In consideration of
this, HM claimed, quoting Vincenti and Kruger\cite{V-K}, that the
correct formula cannot be derived on the basis of
the mean free path theory alone.
They speculated that the correct formula can only be derived using the
Boltzmann equation with the Coriolis force taken into account.

In 2004, however, Clarke and Pringle\cite{C-P} (hereinafter referred to as
CP) showed that the correct viscosity formula for a rotating gas, Eq.~(\ref%
{circular}), can indeed be derived on the basis of
 the mean free path theory, if
one properly takes into account the thermal motion/spread of the molecules.
They obtained the correct result in the inertial frame, thus obviating
inclusion of the Coriolis force. They approximate molecular orbits by
straight lines, which is acceptable under the assumption $\lambda \ll R$, or,
more precisely, under the assumption $\lambda \ll H\ll R$, where $H$ and $R$
are the disc thickness and the radial distance. Note that $H=c/\Omega $,
where $c$ is a typical molecular speed.

Developing CP's idea further, Matsuda and Hayashi\cite{M-H} (hereinafter
referred to as MH) showed that the results can be grasped
more easily and calculated more
simply when calculations are made in a rotating frame. Quoting
MH's conclusion:

\begin{quote}
``... of the molecules originally present in the inner annulus, those whose
angular momenta are larger than the average angular momentum of the outer
annulus will preferentially be transported from the inner to the outer
annulus. The contribution from those molecules that have smaller angular
momenta and have reached the outer annulus is small. Conversely, of the
molecules originally present in the outer annulus, those having angular
momenta smaller than the average angular momentum of the inner annulus will
preferentially be transported to the inner annulus. Accordingly, angular
momentum indeed does flow against its gradients.''
\end{quote}

\begin{figure}[tbp]
\centering
\includegraphics[height=11cm]{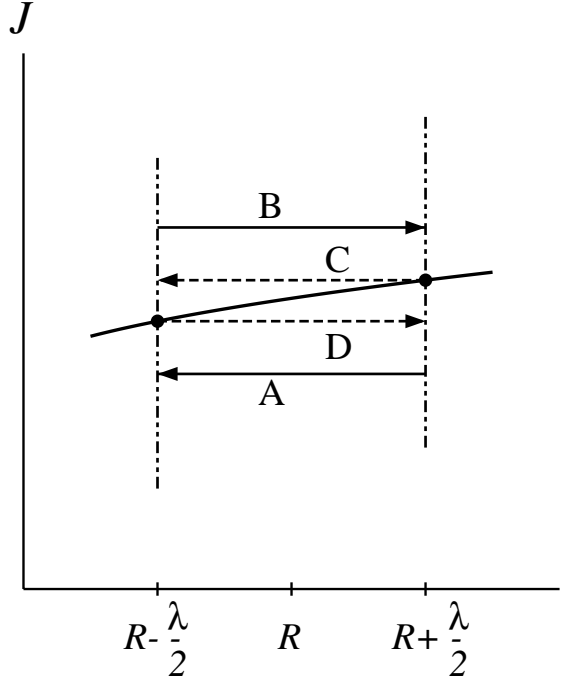}
\caption{Schematic diagram showing the 
angular momentum transport in a Keplerian
disc. The horizontal axis $R$ represents 
the radial coordinate, and the vertical
axis $J$ the specific angular momentum of gas. Molecules or fluid parcels
move from the inner anulus at $R-\protect\lambda /2$ to the outer one at $R+%
\protect\lambda /2$, where $\protect\lambda $ is the mean free path. The
arrows A and B indicate the correct inward/outward angular momentum transport,
while C and D represent
the wrong ones adopted in a textbook. The vertical dotted lines
represent the thermal spread of the angular momentum of molecules/parcels. }
\label{fig1}
\end{figure}

This situation is best described by the schematic diagram in Fig. 1, in
which the horizontal and the vertical axes
represent the radial position $R$ and
the specific angular momentum of the rotating gas $J$ in a Keplerian disc,
respectively. Molecules/fluid parcels in the inner annulus at $R-\lambda /2$
fly to the outer annulus at $R+\lambda /2$ while maintaining their angular
momentum, and vice versa. The solid curve represents the radial distribution of
the angular momentum of the gas. The arrows C and D represent
the angular momentum transport considered by, say, Hartmann\cite{H}. In this
picture, it is apparent that the angular momentum transport is inward. In
order to have the correct picture, we must take into account the thermal
spread of the angular momentum, which is represented by the vertical dotted
lines in Fig. 1. If one calculates the mean angular momentum transfer
correctly, it should be represented by the arrows A and B. In this picture the
angular momentum is transported outward. The original claim of HM, that the
correct formula (1) cannot be derived from the mean free path theory, 
turns out
to be incorrect.

In the SPB paper, it is also claimed that the correct
formula for the viscous torque (1) can be derived on the basis of the mean free
path theory. According to SPB, the correct formula can be derived by
properly taking into account the epicyclic motion of gas parcels. Because
both CP and MH approximate the molecule trajectory as linear,
their arguments are
valid only for $\lambda \ll H$, while SPB treat the case $\lambda \sim H$.
However, we find that the process of deriving the \textquotedblleft
correct\textquotedblright\ formula employed by SPB is incorrect. In \S
2 we briefly review SPB and point out their three major errors. A summary
of our conclusions is given in \S 3.

\section{Process employed by SPB and their errors}

\subsection{Their first error: biased choice of the desired formula}

SPB start from Fig. 2, which, together with its caption in quotation
marks (but excluding the dotted arcs, which were added by the present
authors), is shown here and may be self-explanatory. A fluid parcel A jumps
from the outer annulus to the inner annulus over the mean free path $\lambda 
$ and deposits its angular momentum, $J_{\mbox{\scriptsize{out}}}$, there. 
Parcel B jumps from
the inner annulus to the outer annulus and deposits its angular momentum, $%
J_{\mbox{\scriptsize{in}}}$, there. 
The values of $J_{\mbox{\scriptsize{out}}}$ 
and $J_{\mbox{\scriptsize{in}}}$ are obtained as follows
(the same equation numbers as in SPB are used here):\ 

\begin{figure}[tbp]
\centering
\includegraphics[height=11cm]{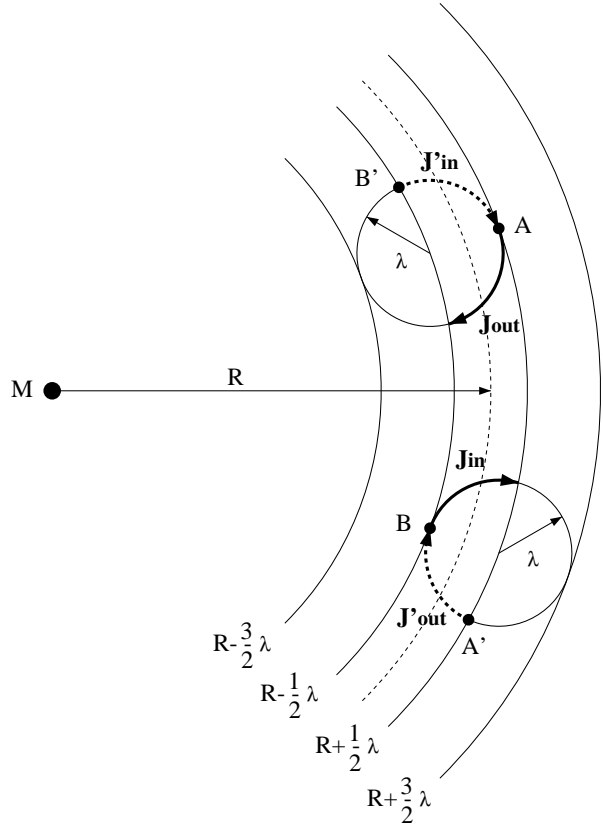}
\caption{``Angular momentum transport in the disc involves the interchange of
two parcels, A and B, initially at radii $R+\protect\lambda /2$ and $R-%
\protect\lambda /2$, respectively, around central mass M. They each
participate in ballistic, epicyclic motion in eddies of radius $\protect%
\lambda $, with A moving inwards and B moving outwards." (Taken
from SPB, with the
radial distance labels rearranged.) The parcel A has a smaller angular
momentum, $J_{\mbox{\scriptsize{out}}}$, 
while B has a larger angular momentum, $J_{\mbox{\scriptsize{in}}}$.
Therefore, one can prove that the angular momentum flows outwards (SPB).
However, there are other possible trajectories of, for example, parcles 
$\mbox{A}^{\prime}$
and $\mbox{B}^{\prime}$ 
having angular momenta $J_{\mbox{\scriptsize{out}}}^{\prime}\,(=
J_{\mbox{\scriptsize{in}}})$ and $%
J_{\mbox{\scriptsize{in}}}^{\prime }\,(=
J_{\mbox{\scriptsize{out}}})$. Thus we could equally prove that the angular
momentum flows inwards. Or, if we take into account A, B,
$\mbox{A}^{\prime}$
and $\mbox{B}^{\prime}$
altogether, we could prove that the angular momentum transfer is zero.
Therefore, this figure alone proves nothing. (See the discussion in \S 2.2.)}
\label{fig2}
\end{figure}

$$
J_{\mbox{\scriptsize{out}}}
=\sqrt{GMR}\left( 1-\frac{1}{4}\frac{\lambda }{R}\right) +O\left[ 
\frac{\lambda ^{2}}{R^{2}}\right] ,\eqno{(23)} 
$$%
$$
J_{\mbox{\scriptsize{in}}}=
\sqrt{GMR}\left( 1+\frac{1}{4}\frac{\lambda }{R}\right) +O\left[ 
\frac{\lambda ^{2}}{R^{2}}\right] .\eqno{(24)} 
$$
The above step, at first glance, seems to be valid, provided that we accept
Fig. 2 (SPB's original, i.e. without dotted arcs). However,
 this is
proved to be incorrect in \S 2.2. Because $J_{\mbox{\scriptsize{in}}}>
J_{\mbox{\scriptsize{out}}}$, they are also to claim that
the angular momentum flows outwards. For the moment, let us accept the above
argument.

The next steps they take, however, are problematic. 
In particular, they derive the
``correct'' formula $J_{\mbox{\scriptsize{in}}}-
J_{\mbox{\scriptsize{out}}} \propto R^2 d\Omega/d R$ as follows
(SPB):

\begin{quote}
``The angular velocity $\Omega(R)$ in a quasi-Keplerian accretion disc is
very close to the Keplerian value, and therefore we can write

$$
\Omega(R) = \sqrt{ \frac{GM}{R^3}}. \eqno{(25)} 
$$
It follows that

$$
R^2 \Omega\left( R+\frac{\lambda}{6} \right ) = \sqrt{GMR} \left( 1 + \frac16%
\frac{\lambda}{R} \right ) ^{-3/2}, \eqno{(26)} 
$$
or, to first order in $\lambda/R$,

$$
R^2 \Omega\left( R+\frac{\lambda}{6} \right ) = \sqrt{GMR} \left( 1 - \frac14%
\frac{\lambda}{R} \right ) + O\left [\frac{\lambda^2}{R^2} \right]. %
\eqno{(27)} 
$$

This also implies that

$$
R^2 \Omega\left( R-\frac{\lambda}{6} \right ) = \sqrt{GMR} \left( 1 + \frac14%
\frac{\lambda}{R} \right ) + O\left [\frac{\lambda^2}{R^2} \right ]. %
\eqno{(28)} 
$$

Comparing equations (23) and (24) with equations (27) and (28), we find that
to first order in $\lambda/R$, the net angular momentum transfer is given by

$$
J_{\mbox{\scriptsize{in}}}-
J_{\mbox{\scriptsize{out}}} = R^2 \Omega\left( R-\frac{\lambda}{6} \right ) -R^2
\Omega\left( R+\frac{\lambda}{6} \right ) \simeq -\frac{\lambda}{3} R^2 
\frac{d\Omega}{dR}. \eqno{(29)} 
$$

(a few lines omitted)\newline
Hence we have demonstrated using a simple heuristic derivation that the
viscous torque is indeed proportional to the gradient of the angular
velocity in an accretion disc within the context of a mean free path,
parcel-exchange picture.''
\end{quote}

\medskip The steps above taken by SPB are incorrect. Firstly, why does $%
\lambda /6$ appear in Eqs. (26)-(29)? 
In fact, this has been chosen
so as to derive $\lambda /4R$ in Eqs. (23) and (24). In the following
we show how the coefficient $1/6$ can be obtained.

Let us assume that $J_{\mbox{\scriptsize{in}}}
-J_{\mbox{\scriptsize{out}}} \propto R^2 (d \Omega/dR)$ and
attempt to obtain a relevant parameter $\alpha$ to multiply $\lambda$. 
First, we write down
the following equation to obtain $\alpha$:

$$
R^2 \Omega(R+\alpha \lambda)= \sqrt{GMR} \left
( 1- \frac14 \frac{\lambda}{R%
} \right ). \eqno{(\mbox{A})} 
$$
Then, assuming $\lambda \ll R$, we can expand the left-hand side and retain the
first term, obtaining

$$
\mbox{LHS}\approx R^{2}(\Omega +\alpha \lambda d\Omega /dR)=R^{2}\left( \Omega
+\alpha \lambda \left( -\frac{3\Omega }{2R}\right) \right) \newline
=\sqrt{GMR}\left( 1-\frac{3\alpha \lambda }{2R}\right) .
\eqno{(\mbox{B})}\newline
$$%
Finally, equating the coefficients of the last terms in (A) and (B), we get $\alpha
=1/6$. This value of $\alpha $, together with that for Eq. (28), leads
to Eq. (29) which they sought.

We now show that with their argument, we could derive any
(invalid) formula we desire. First, let us assume the
following (invalid) relation: 
$J_{\mbox{\scriptsize{in}}}-J_{\mbox{\scriptsize{out}}}
 \propto Rd v_{\phi}/dR $, where $%
v_{\phi}=R \Omega$ is the azimuthal velocity. With this, we
can prove the following
relation:

$$
Rv_{\phi }(R+\frac{1}{2}\lambda )=\sqrt{GMR}\left( 1-\frac{1}{4}\frac{%
\lambda }{R}\right) .\eqno{(27)'} 
$$%
\noindent The numerical coefficient $1/2$ multiplying $\lambda $ 
on the LHS of
$(27)^{\prime}$ is obtained by solving the following equation for $\beta $:

$$
R(R+\beta\lambda) \Omega(R+\beta \lambda)= \sqrt{GMR} \left
( 1- \frac14 
\frac{\lambda}{R} \right ). \eqno{(\mbox{C})} 
$$
\noindent To obtain $\beta=1/2$ is straightforward.

The above argument clearly shows that their result, (29), is not unique, as
it demonstrates that we can also derive 
$J_{\mbox{\scriptsize{in}}}-J_{\mbox{\scriptsize{out}}}\propto Rdv_{\phi }/dR$,
which is obviously incorrect. Indeed, this derivation suggests that
we can derive further similar relations. We may in fact assume the following
general formula:

$$
J_{\mbox{\scriptsize{in}}}-
J_{\mbox{\scriptsize{out}}}\propto R^{n}d(R^{2-n}\Omega )/dR\propto
R^{n}d(R^{1/2-n})/dR,\qquad 1/2<n.\eqno{(\mbox{D})} 
$$
This general formula indicates that there are an infinite number of
formulae satisfying $1/2<n$. One such \textquotedblleft
false\textquotedblright\ formula, 
$J_{\mbox{\scriptsize{in}}}-J_{\mbox{\scriptsize{out}}}
\propto Rd(R\Omega )/dR$,
which we presented above, corresponds to the case $n=1$. The correct formula
corresponds to the case $n=2$.

SPB's argument is similar to deriving an equation from a solution, a
procedure which is generally invalid.

\subsection{Second error in SPB: biased choice of parcel trajectories}

Secondly, in Fig. 2, the thick solid arcs starting at A and B are not the
only possible trajectories. We accept SPB's approximation of the epicycle as
circular, (although, in fact, it should be an ellipse having a major axis to
minor axis ratio of 2:1 if the central gravity is taken into account). In
Fig. 2 we add examples of other possible trajectories of parcels $\mbox{A}^
{\prime}$
and $\mbox{B}^{\prime}$ in dotted arcs. As apparent from the figure,
$\mbox{A}^{\prime}$
and $\mbox{B}^{\prime}$ start from
the outer and inner annulus, respectively, carrying their angular momenta $%
J_{\mbox{\scriptsize{out}}}^{\prime }$ and 
$J_{\mbox{\scriptsize{in}}}^{\prime }$. In this case 
$J_{\mbox{\scriptsize{out}}}^{\prime
}\,(=J_{\mbox{\scriptsize{in}}})$ is larger than 
$J_{\mbox{\scriptsize{in}}}^{\prime }\,(=J_{\mbox{\scriptsize{out}}})$, 
just contrary to $%
J_{\mbox{\scriptsize{out}}}<J_{\mbox{\scriptsize{in}}}$ derived in SPB. 
Thus we can similarly prove that the angular
momentum flows inwards rather than outwards. Or, if we take into account A,
B, $\mbox{A}^
{\prime}$
and $\mbox{B}^{\prime}$ altogether, we could prove that the angular momentum transport
is zero. The parcel trajectories chosen by SPB are thus just those
favorable to their argument. Figure 2 alone does not prove that the angular
momentum flows outwards.

In order to obtain the correct conclusion, that the angular momentum flows
outwards, one has to take into account the velocity distribution or
thermal spread of molecules at the ejecting point. Then we find that
particles ejected in the forward direction, carrying larger angular momenta
than those carried by the mean flow at the ejection point, are
preferentially transported outwards. This was shown by MH in the limit of
a short mean free path. The same consideration should also apply to the case
of a larger mean free path.

We now raise another question with regard to Fig. 2, which was originally
presented by SPB. Let us specify the molecular trajectories,
 rather than a vague
notion of the parcel motion of turbulent eddies. In kinetic theory, the
mean free path, $\lambda $, and the radius of the epicycle, $\sim H$, are
completely different concepts. The former is defined as $\lambda =1/n\sigma $%
, where $n$ and $\sigma $ are the number density and the cross section of
the molecules, respectively. Contrastingly, $H$ is defined as $c/\Omega $%
, where $c$ is the thermal molecular velocity. SPB confuses these two
different concepts.

\subsection{Third error in SPB: reference frames}

Let us comment on a criticism of HM made by SPB. In their \S 3.3,
where they discuss the derivation of the viscous torque by FKR, they write
(here again the equation numbers are theirs) the following:

\begin{quote}
``... They (HM) assert that the linear velocity of the plasma at $R -
\lambda/2$ as viewed by an observer at radius R should instead be given by

$$
v_{\mbox{\scriptsize{rel}}} \left 
( R - \frac{\lambda}{2} \right ) = \left ( R - \frac{%
\lambda}{2} \right ) \Omega\left ( R - \frac{\lambda}{2} \right )
-R\Omega(R) + \Omega(R)\frac{\lambda}{2}. \eqno{(13)} 
$$

By employing this expression for $v_{\mbox{\scriptsize{rel}}}$ 
and following the same
procedure used to obtain equation (12), they find that to first order in $%
\lambda$,

$$
{\mathcal{G}} = L_{\mbox{\scriptsize{in}}}-
L_{\mbox{\scriptsize{out}}} = -2 \pi R^3 \beta\nu \Sigma \frac{d\Omega} {%
dR}. \eqno{(14)} 
$$

This expression for $\mathcal{G}$ does indeed have the correct dependence on 
$d\Omega /dR$, but nonetheless we claim that equation (13) for the relative
velocity used by HM is incorrect. The correct expressions for the relative
velocities are in fact (see, e.g., Mihalas \& Binney 1981)

\[
v_{\mbox{\scriptsize{rel}}} \left ( R - \frac{\lambda}{2} \right ) 
= \left ( R - \frac{%
\lambda}{2} \right ) \Omega\left ( R - \frac{\lambda}{2} \right )
-R\Omega(R) 
\]
$$
v_{\mbox{\scriptsize{rel}}} 
\left ( R + \frac{\lambda}{2} \right ) = \left ( R + \frac{%
\lambda}{2} \right ) \Omega\left ( R + \frac{\lambda}{2} \right )
-R\Omega(R), \eqno{(15)} 
$$
where $v_{\mbox{\scriptsize{rel}}}(R-\lambda/2) $ 
denotes the velocity of a plasma parcel
at $R-\lambda/2$ as seen by an observer at $R$, and 
$v_{\mbox{\scriptsize{rel}}} (R+
\lambda/2)$ denotes the velocity of a plasma parcel at $R+\lambda/2$ as seen
by an observer at $R$. ...''
\end{quote}

\medskip We point out here that HM's choice of the relative velocity in
Eq. (13) is based on the (co-moving) rotating frame at
radial position $R$. By contrast,
SPB's choice of Eq. (15) implies that the relative velocity is
observed in the (co-moving) non-rotating frame.
Therefore, it is not a matter of \textquotedblleft
correct\textquotedblright\ or \textquotedblleft incorrect\textquotedblright
. As explained in MH (p. 360), any frame can and should, with the succeeding
calculations performed correctly, yield the same result.
There are four typical frames:

\begin{enumerate}
\item The rest frame, which is an inertial frame,
 with respect to the central mass. This frame was employed
by Hartmann\cite{H}. In this frame there is conservation of angular momentum.
 The angular momentum computed in this frame is called
the absolute angular momentum, which is conserved. The velocity of fluid at $%
R-\lambda /2$ is observed as $(R-\lambda /2)\Omega (R-\lambda /2)$.

\item The rotating frame, in which an observer is rotating with
the angular velocity $\Omega$ in the above rest frame.  This frame 
was employed by FKR, and
the fluid velocity is seen as $(R-\lambda/2)\Omega(R-\lambda/2)+\Omega(R)
\lambda/2$.

\item The co-moving non-rotating frame employed by SPB. 

\item The (co-moving) rotating frame (see Eq. (13)) suggested by HM for
modification of the approach of
 FKR in order to obtain a seemingly correct result.
\end{enumerate}

\medskip In any frame, if one does not properly take into account the
thermal motion of molecules, but otherwise carries out the calculation 
correctly, one
inevitably reaches the incorrect conclusion that the viscous stress is
proportional to $d(R^{2}\Omega )/dR$. Here, by ``correctly" we mean that the
convected variable is taken as the absolute angular momentum.
On the other hand, if one calculates the angular momentum using the apparent
fluid velocity, $v_{\mbox{\scriptsize{rel}}}$, 
one reaches the conclusion that the
viscous stress is proportional to $d(R\Omega )/dR$ in frames 2 and 3 and 
to $d\Omega /dR$ in frame 4. The last result seems to be the correct one.
However, this conclusion is derived on the basis of
 the invalid assumption that the
``apparent" angular momentum is conserved. Only the absolute angular momentum
is conserved. We therefore cannot claim 
that such a simple manipulation leads to
the correct answer. This is the point made by HM, and it is misinterpreted by
SPB in the above quotation.

\section{Conclusions}

\begin{enumerate}
\item \medskip \noindent\ Subramanian, Pujari and Becker\cite{S-P-B} (SPB)
claim to have derived a formula according to which
 the angular momentum transport is
proportional to $R^{2}d\Omega /dR$, where $R$ and $\Omega $ are the radial
distance and the angular velocity of the rotating disc, respectively.
However, their derivation assumes, as its starting point, the formula
to be derived, and it is therefore ill-founded from a mathematical point
of view.
Following their procedure, we could also derive, for example, a 
(invalid) formula according to which the angular momentum transport 
is proportional to $%
Rdv_{\phi }/dR$, where $v_{\phi }$ is the azimuthal velocity.

\item \medskip\ They chose parcel (molecule) trajectories favorable to their
desired result. We could in the same way prove that the angular momentum flows
inwards by choosing another set of trajectories. Moreover, they confused the
concepts of the mean free path and the radius of an epicycle.

\item \medskip\ Their criticism of the formula for the relative velocity
given in our previous paper (HM) is irrelevant. We chose a rotating frame,
while they chose a non-rotating frame. This choice is not a matter of 
correctness or incorrectness. 
We studied four typical frames in the present paper.
In any frame, if one carries out the calculation properly but ignores
the thermal spread/motion of molecules, one inevitably ends up with 
the incorrect
conclusion that the viscous stress is proportional to $d(R^{2}\Omega )/dR$.

\end{enumerate}

\section*{Acknowledgements}

The authors would like to thank anonymous referees for helpful 
and useful advice and suggestions, and also thank Dr. H. M. J. Boffin 
for his careful reading of
the manuscript.  
T. M. is supported by a Grant-in-Aid for Scientific Research
of the Japan Society for the Promotion of Science (13640241) and by
\textquotedblleft The 21st Century COE Program of Origin and Evolution of
Planetary Systems" of the Ministry of Education, Culture, Sports, Science
and Technology (MEXT).

\end{document}